\documentclass[10pt,conference]{IEEEtran}
\usepackage{cite}
\usepackage{amsmath,amssymb,amsfonts}
\usepackage{algorithmic}
\usepackage{graphicx}
\usepackage{textcomp}
\usepackage{xcolor}
\usepackage[hyphens]{url}

\usepackage{algorithm}
\usepackage{caption}
\usepackage{subcaption}
\usepackage{authblk}
\usepackage[english]{babel}
\usepackage{amsthm}
\usepackage{bm}
\usepackage{svg}
\usepackage{float}

\def\BibTeX{{\rm B\kern-.05em{\sc i\kern-.025em b}\kern-.08em
    T\kern-.1667em\lower.7ex\hbox{E}\kern-.125emX}}

\title{Wide Quantum Circuit Optimization with Topology Aware Synthesis}
\author[1]{Mathias Weiden}
\author[2]{Ed Younis}
\author[1]{Justin Kalloor}
\author[1]{John Kubiatowicz}
\author[2]{Costin Iancu}
\affil[1]{Department of Electrical Engineering and Computer Science, University of California, Berkeley}
\affil[ ]{\textit{\{mtweiden, jkalloor3, kubitron\}@cs.berkeley.edu}}
\affil[2]{Computational Research Division, Lawrence Berkeley National Laboratory}
\affil[ ]{\textit{\{edyounis, cciancu\}@lbl.gov}}

\title{Improving Quantum Circuit Synthesis with \ Machine Learning}
\newtheorem{definition}{Definition}[section]
\begin{document}
\maketitle
\thispagestyle{plain}
\pagestyle{plain}


\begin{abstract}
In the Noisy Intermediate Scale Quantum (NISQ) era, finding implementations of quantum algorithms that minimize the number of expensive and error prone multi-qubit gates is vital to ensure computations produce meaningful outputs. Unitary synthesis, the process of finding a quantum circuit that implements some target unitary matrix, is able to solve this problem optimally in many cases. However, current bottom-up unitary synthesis algorithms are limited by their exponentially growing run times. We show how applying machine learning to unitary datasets permits drastic speedups for synthesis algorithms. This paper presents QSeed, a seeded synthesis algorithm that employs a learned model to quickly propose resource efficient circuit implementations of unitaries. QSeed maintains low gate counts and offers a speedup of $3.7\times$ in synthesis time over the state of the art for a 64 qubit modular exponentiation circuit, a core component in Shor's factoring algorithm. QSeed's performance improvements also generalize to families of circuits not seen during the training process.
\end{abstract}

\section{Introduction}
\label{section:introduction}
Quantum computers are capable of solving many problems that are thought to be intractable on classical computers \cite{nielsen2002quantum}. In the Noisy Intermediate Scale Quantum (NISQ) \cite{preskill_2018_nisq} era, where full error correction is not yet realizable, careful consideration of error mitigation techniques in compilation are vitally important. To this end, techniques such as random compilation \cite{Wallman_2016}, noise-adaptive mapping \cite{Tannu_2018, Murali_2019}, and zero noise extrapolation \cite{Giurgica_2020_zeronoise} have been proposed. Google's quantum supremacy experiments demonstrated empirically that simply reducing the number of operations remains an impactful noise mitigation strategy \cite{arute2019supremacy}. The need for optimization algorithms that reduce gate counts is therefore critical.

Rule based circuit optimization algorithms are widely used because they are quick and effective for small circuits \cite{Prasad_2006, Amy_2013, Kissinger_2019, Heyfron_2017}. These techniques rely on recognizing small gate patterns and replacing them with more efficient implementations, and by making small, heuristic-guided transformations. Recently, deep Reinforcement Learning has emerged as a technique to automate the discovery and application of circuit transformation rules \cite{Fosel_2021, Quetschlich_2022, Ostaszewski_2021}. However, these techniques remain hindered by a narrow, local view of the circuits they optimize or by the limitations of human intuited heuristics.

Unitary synthesis is a bottom-up method for constructing circuit implementations of unitary matrices. It has become popular as an optimization technique as it eschews a gate level view of the program for a more global perspective. This broader frame of reference allows for optimization that is otherwise impossible with rule based methods. State-of-the-art synthesis algorithms have demonstrated that they are capable of producing the best known implementations of a variety of circuits \cite{davis_heuristics_2019, smith2021leap, younis_2020_qfast}. Synthesis searches over a tree of increasingly more complex candidate circuits and evaluates each one's ability to implement a unitary target. Yet, these techniques remain limited in their applicability due to their exponentially growing run times. Circuits wider than five qubits must be partitioned into smaller subcircuits, which can each be individually synthesized. Importantly, even in these cases, where synthesis is not given a full global view of the circuit being optimized, it still provides unmatched optimization results \cite{wu2021qgo, weiden_2022_topas}.

The core contribution of this paper is a method for replacing the expensive circuit tree search of unitary synthesis with Machine Learning (ML) inference. Given a unitary matrix, we immediately propose resource efficient circuit templates that, with high probability, are able to directly implement the unitary. We present our findings relating to the learnability of certain sets of unitary matrices. To our knowledge, we are the first to apply ML to unitary datasets for optimization. We believe that our approach is generic and will apply to the field of quantum compilation no matter how quantum programming evolves. This project focuses on demonstrating the advantages of leveraging ML for the synthesis of three qubit unitaries, but remains both applicable and promising for accelerating synthesis of larger matrices. With the use of circuit partitioning, we demonstrate how QSeed is capable of scaling to circuits of 100 qubits and beyond. Our technique provides substantial speedups compared to state-of-the-art synthesis algorithms, ensuring more practical applicability of this style of optimization. 

The paper is divided as follows: Section \ref{section:background} describes the unitary synthesis process and illustrates its scalability limitations. Section \ref{section:learning} discusses our findings on applying ML datasets of unitaries. Section \ref{section:qseed} introduces QSeed, a seeded synthesis algorithm that uses ML to accelerate the synthesis tree search process. Section \ref{section:evaluation} evaluates QSeed's performance. Section \ref{section:discussion} provides further discussion related to QSeed's performance and future research directions.

\section{Background}
\label{section:background}
In this section we present necessary context and provide definitions for notable concepts.

\subsection{Unitary Synthesis and Parameterized Quantum Circuits}
Unitary synthesis algorithms are a class of algorithms that take as input a unitary matrix and iteratively build up a parameterizable quantum circuit implementing that unitary. 
\begin{definition}[Unitary Synthesis]
    \label{def:synthesis}
    Given a target unitary matrix $U_T \in \mathbb{U}(N)$, a unitary synthesis algorithm constructs a quantum circuit that implements a unitary $U_S$ satisfying
    \[
        \| U_T - U_S \|_{HS} \leq \epsilon
    \]
    where $\|\cdot\|_{HS}$ is the Hilbert-Schmidt norm defined as
    \[
        \|A \|_{HS} = Tr(A^\dagger A).
    \]
\end{definition}
\begin{definition}[Parameterized Quantum Circuit]
    A Parameterized Quantum Circuit (PQC) is a quantum circuit that contains gates whose operations are a function of real valued parameters. 
\end{definition}

\begin{figure}[b]
    \centering
    \includegraphics[width=0.48\textwidth]{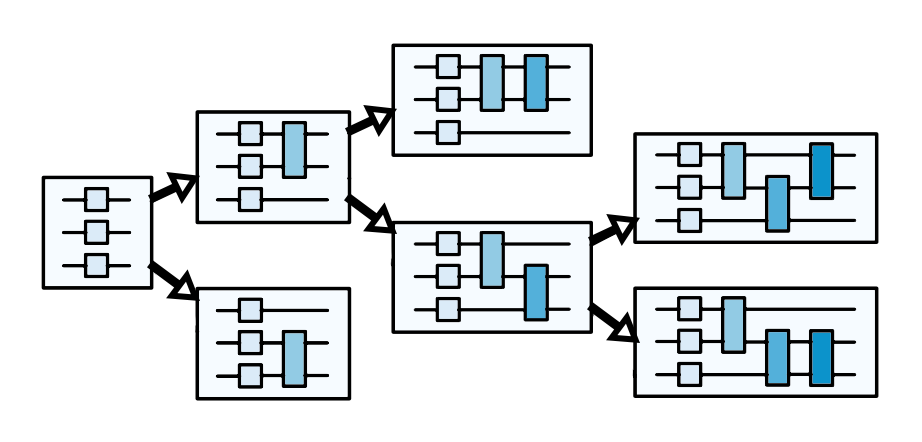}
    \caption{In bottom-up synthesis, parameterizable sequences of gates are appended to a base PQC. Circuits grow according to a circuit tree where parent circuits serve as prefixes to child circuits.}
    \label{fig:qsearch_tree}
\end{figure}

Unitary synthesis algorithms such as QSearch \cite{davis_heuristics_2019} and LEAP \cite{smith2021leap} are capable of finding the best known implementations of a variety of quantum algorithms. In these synthesis algorithms, sequences of parameterizable multi-qubit gates are appended to a simple base PQC. This style of synthesis is referred to as bottom-up synthesis and is depicted in Fig.~\ref{fig:qsearch_tree}. The allowed placement of these gates is determined by a \emph{qubit topology}. 

\begin{definition}[Qubit Topology]
    A qubit topology is a graph $G = (V,E)$. Vertices $v \in V$ represent qubits, while edges $(u,v) \in E$ represent (hardware) allowed interactions between qubits.
\end{definition}

The bottom-up synthesis process can be viewed as a search through a tree of parameterizable circuits. Traversal begins at the tree's root node, and child nodes are explored until a circuit capable of implementing the target unitary is found. When visiting a node in this traversal, the parameters in the associated circuit must be solved for in order to minimize the distance described in Definition \ref{def:synthesis}. This process of solving for parameters is called \emph{instantiation}. This computationally expensive process can be thought of as deciding whether or not a PQC is suitable for the implementation of a unitary.

\begin{definition}[Instantiation]
    Given a target unitary $U_T$ and a PQC represented by a unitary operator $U(\bm{\theta})$ parameterized by $\bm{\theta} \in \mathbb{R}^m$, instantiation solves the problem \[\arg\min_{\bm{\theta}} \|U_T - U(\bm{\theta})\|_{HS}.\]
\end{definition}

\subsection{Run Time Expense}
The run time of unitary synthesis algorithms such as QSearch and LEAP grows exponentially in the number of qubits. This is due in part to the fact that the size of the unitaries that must be evaluated during instantiation grows as $O(2^n)$. Furthermore, the maximum number of CNOT gates needed to implement an arbitrary unitary is $\frac{1}{4} \lceil 4^n - 3n - 1 \rceil$ \cite{shende2006synthesis}. This means the synthesis circuit tree's depth grows exponentially. As the size of the tree grows, synthesis algorithms also tend to explore regions that do not contain solution circuits. In these scenarios, synthesis must backtrack up the tree and probe new regions. These things combined mean the amount of time spent searching the synthesis circuit tree quickly becomes untenable. 

\subsection{Circuit Partitioning}
\begin{figure}[b]
    \centering
    \includegraphics[width=0.3\textwidth]{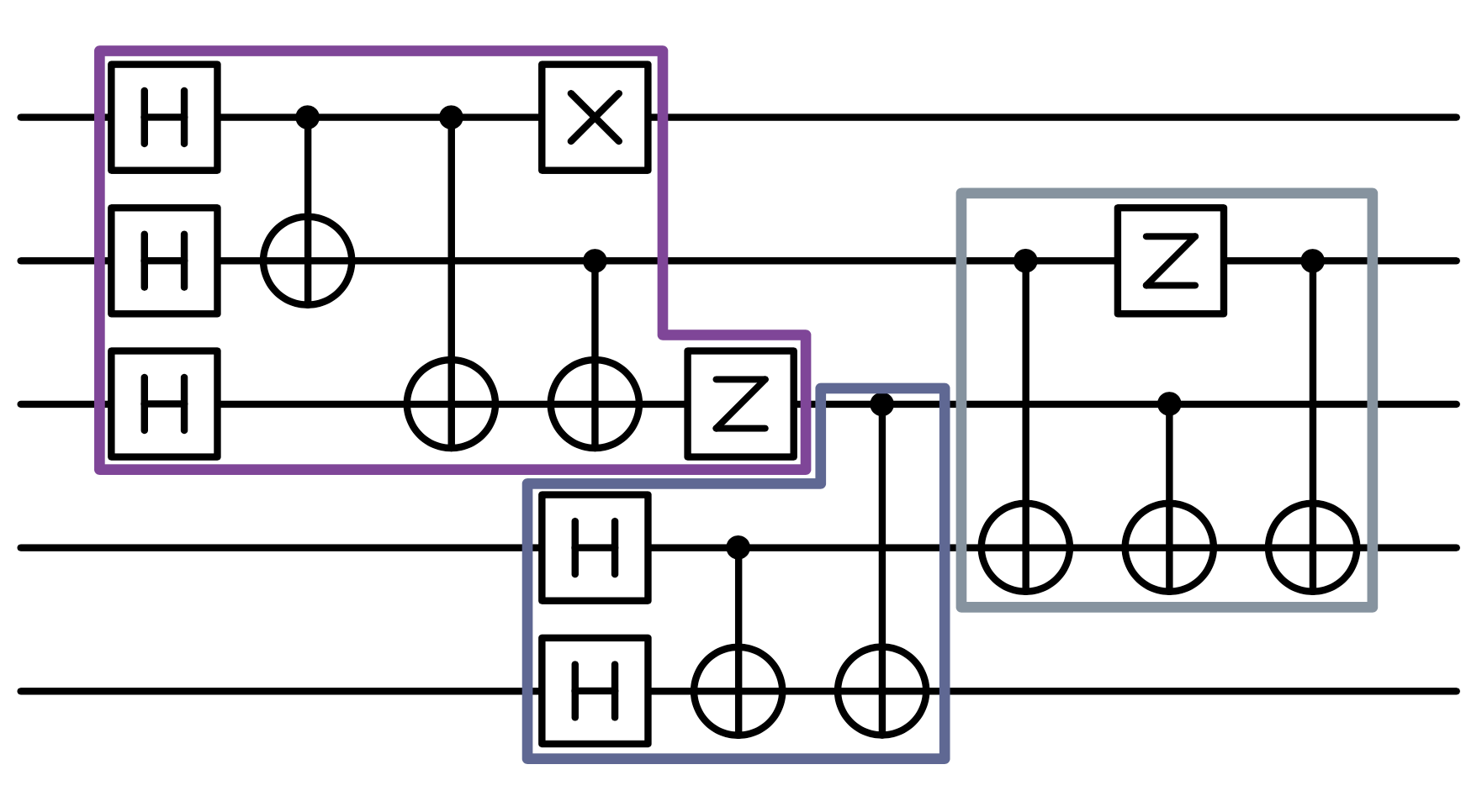}
    \caption{A width five quantum circuit that has been partitioned into subcircuits of three qubits.}
    \label{fig:partitioned_circuit}
\end{figure}

The exponentially scaling run time of bottom-up synthesis algorithms limits the width of, or number of qubits in, circuits that can be synthesized. Wide quantum circuits must be split up, or \emph{partitioned}, into subcircuits that can be synthesized independently. Doing so enables synthesis to tackle optimization problems up to 100 qubits and beyond. An example of a partitioned circuit can be seen in Fig.~\ref{fig:partitioned_circuit}. The partitioning strategy used is described in the QUEST and TopAS projects \cite{patel_2022_quest, weiden_2022_topas}.

By partitioning, the synthesis of a single large quantum circuit is broken into many smaller pieces that can be handled individually and in parallel. The run time of synthesis algorithms applied to partitioned circuits is no longer limited by the exponentially growing cost of unitary function evaluation, but by the number of partitions formed. This number scales polynomially with a circuit's depth and width. Even so, with large circuits, the number of partitions formed can be high enough to cause intractability, even without this exponentially scaling cost.

\section{Learning on Unitary Matrices}
\label{section:learning}
Here, we present our findings relating to the learnability of datasets which consist of unitary matrices. Say we have $n$ qubits, and a set of unitaries $\mathcal{U}(2^n)$ that represent operations we can perform on our qubits. For our $n$ qubits, we can also construct a set that contains all parameterized quantum circuits with $K$ or fewer multi-qubit gates. We call this set $\mathcal{PQC}(n, K)$.

What we wish to do is find a mapping 
\[
    \phi : \mathcal{U}(2^n) \longrightarrow \mathcal{PQC}(n,K)
\]
such that each unitary $U \in \mathcal{U}(2^n)$ is mapped to the circuit $PQC \in \mathcal{PQC}(n,K)$ that implements $U$ with the fewest number of multi-qubit gates. We assume that $K$ is large enough so that some circuit in $\mathcal{PQC}(n,K)$ implements $U$ to within some error $\epsilon$. In this section, we describe the efforts we have made to solve this problem by training a deep neural network to learn this mapping $\phi$.

\subsection{Canonical Representation of Unitaries}
\label{section:canonical}
It is a well known fact of quantum mechanics that two unitaries that differ only by a global phase will have identical measurement statistics \cite{nielsen2002quantum}. For the sake of easing the training of a neural network, it is desirable to have a representation of unitary matrices such that any two unitaries that differ only by a global phase are represented the same way. This way unitaries that are, for all intents and purposes, the same, are represented identically. Formally, we want a function $f$ such that for unitaries $U, U^*$,
\[
    \forall \theta \in \mathbb{R} \quad U = e^{i\theta} U^* \implies f (U) = U^* .
\]
If $f(U) = U^*$ we call the matrix $U^*$ the canonical representation of $U$.

The process of computing $f(U) = U^*$ happens in two steps:
\begin{enumerate}
    \item Convert $U$ to a special unitary $V\in \mathbb{SU}(N)$ so that \[V = e^{i 2\pi k / N}U^*\] for some integer $k$.
    \item Find the integer $k$ and multiply \[U^* = e^{-i 2 \pi k /N} V\] so that the complex phase of the first nonzero element of $V$ is cancelled.
\end{enumerate}
Performing these two steps ensures that each unitary is mapped to a canonical special unitary.

\begin{proof}
    The first step ensures that the determinant $|V| = 1$. However, each $U \in \mathbb{U}(N)$ could be mapped to any one of the $N$ possible special unitaries in the form $V = e^{i 2 \pi k / N} U^\star$ for $k \in \mathbb{Z}$. To ensure that $f(\cdot)$ yields the same special unitary no matter the global phase, we perform the second step. Even with different global phases, two special unitaries $V_1 = e^{i 2 \pi k_1 / N} U^\star$ and $ V_2 = e^{i 2 \pi k_2 / N} U^\star$ that differ only by some root of unity global phase will have nonzero elements in the same locations. Because $V_1, V_2$ are unitary, there must be at least one nonzero element in each of their rows and columns. Ensuring that the complex phase of the first elements of $V_1, V_2$ are the both zero will yield the same total global phase. Thus, $f(V_1) = f(V_2) = U^*$.
\end{proof}

\subsection{Dimensionality of Unitary Datasets}
\label{section:low_dim}
Various works in the fields of Variational Quantum Algorithms (VQAs) and Quantum Machine Learning (QML) have demonstrated the high expressive power of parameterized circuits \cite{harrow_2009_random, sim_2019_expressibility, du_2020_expressivepower, holmes_2022_expressibility}. A single PQC can express many unitaries, and one unitary is likely to be implementable by many PQCs. This makes the problem of learning a mapping from unitaries to circuits challenging. 

\begin{figure}[]
    \centering
    \includegraphics[width=0.48\textwidth]{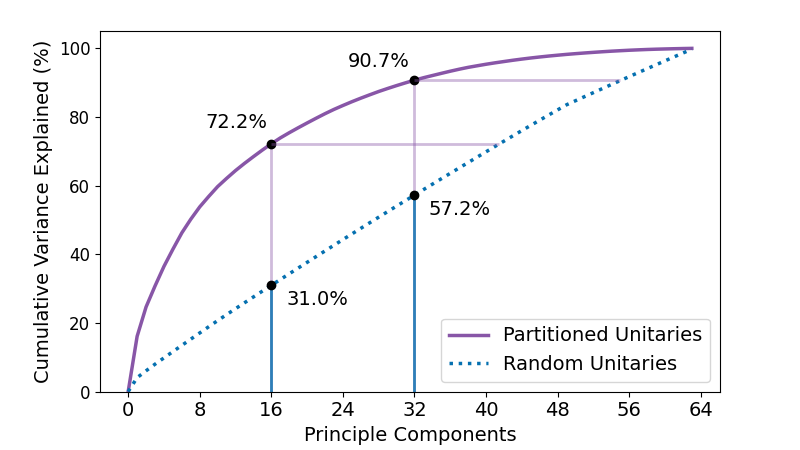}
    \caption{Cumulative variance explained by Principle Components of two datasets of three qubit unitaries. One is randomly generated, the other is made of partitioned unitaries from circuits in Table \ref{tab:test_circuits}. The PCA of the partitioned unitaries implies these unitaries lie on a low dimensional manifold; there are patterns that can be identified in this dataset. The random unitaries do not exhibit identifiable low dimensional patterns. Learning on random unitaries is difficult, learning on partitioned unitaries is easier.}
    \label{fig:pca}
\end{figure}

To better convey how this issue relates to the learnability of unitary datasets, we can examine a set of random unitaries generated by PQCs. Doing so allows us to make observations about the underlying dimensionality of the unitaries that a set of PQCs can implement. This is helpful because data that exhibit some form of low dimensionality are more easily learned by neural networks \cite{rumelhart_1986_internal_representations}. To quantify dimensionality, we can perform a Principle Component Analysis (PCA) \cite{jolliffe_2016_pca} on this set of unitaries. Fig.~\ref{fig:pca} illustrates the variance explained by the principle components of a set of $2^3 \times 2^3$ complex unitaries randomly generated by twelve PQCs. As the variance in this dataset is spread evenly across each principle component, we have little expectation that we can learn a generalizable mapping from unitaries to PQCs for this dataset.

In practice, we need only learn a mapping from unitaries to PQCs for unitaries and PQCs of interest. Consider for example the task of optimizing wide quantum circuits taken from a diverse benchmark suite. If we partition the circuits outlined in Table \ref{tab:test_circuits} into unitaries of three qubits as mentioned in Section \ref{section:background}, we obtain a dataset which is far lower dimensional than in the randomly generated unitary case. Fig.~\ref{fig:pca} illustrates how linear combinations of the first 16 principle components of this partitioned unitary dataset describe 72.2\% of the data's variance.

Because unitaries taken from partitioned quantum circuits exhibit patterns that lie on low dimensional manifolds, we expect that learning a mapping from these unitaries of interest to the PQCs that can express them will be possible.

\begin{figure}
    \centering
    \begin{subfigure}[b]{0.20\textwidth}
        \includegraphics[width=\textwidth]{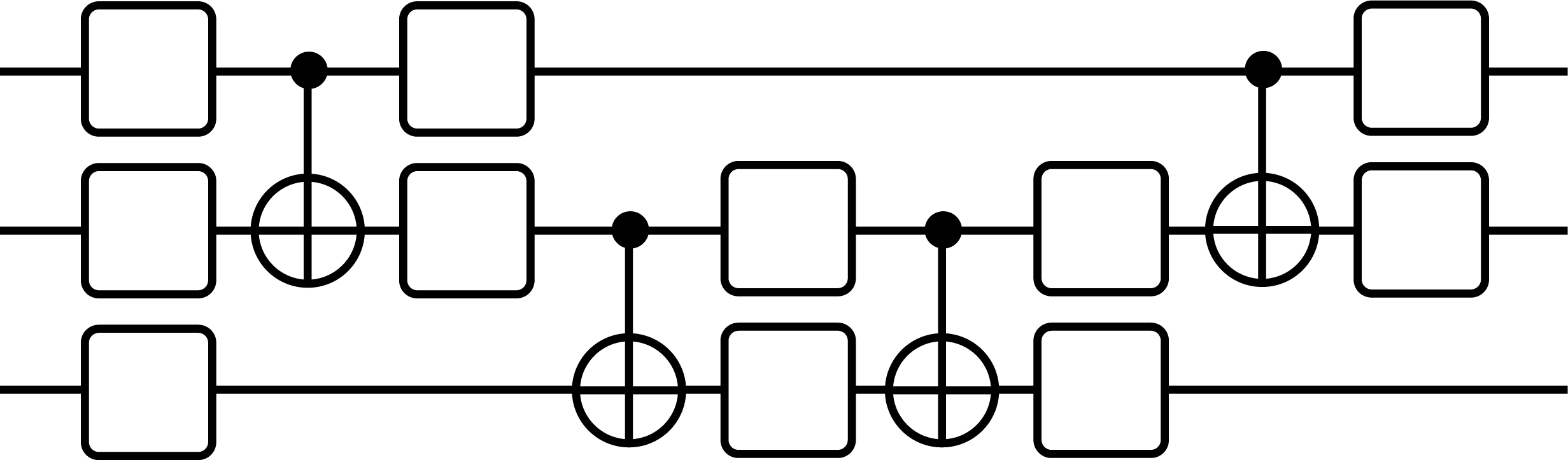}
        \caption{Template B}
        \label{fig:template_b}
    \end{subfigure}
    \begin{subfigure}[b]{0.20\textwidth}
        \includegraphics[width=\textwidth]{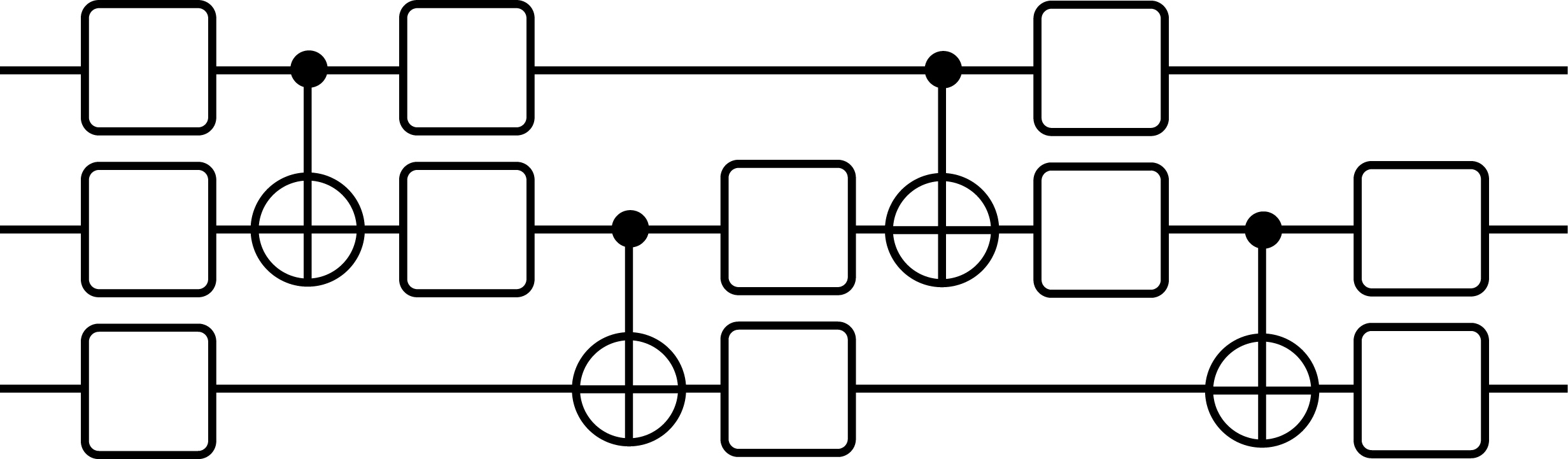}
        \caption{Template C}
    \end{subfigure}
    \par\bigskip
    \begin{subfigure}[b]{0.28\textwidth}
        \includegraphics[width=\textwidth]{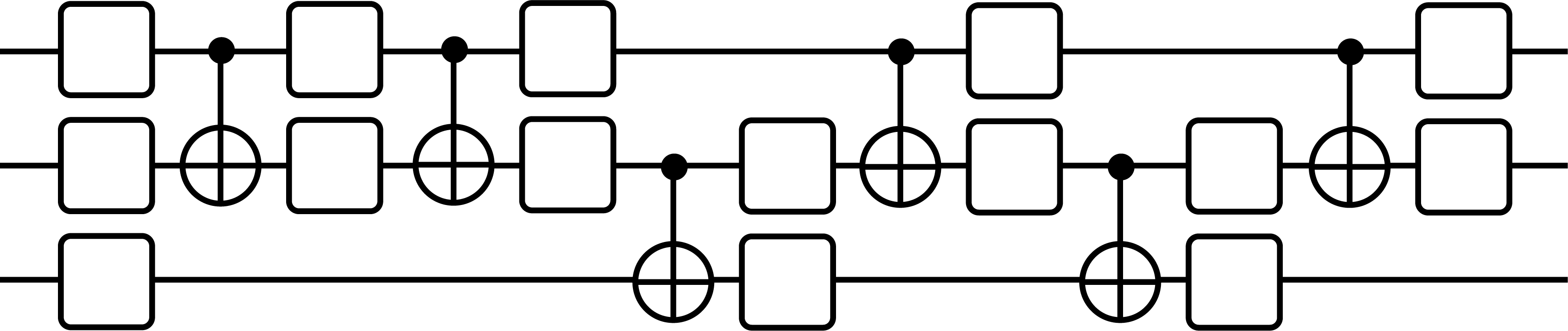}
        \caption{Template I}
    \end{subfigure}
    \par\bigskip
    \begin{subfigure}[b]{0.35\textwidth}
        \includegraphics[width=\textwidth]{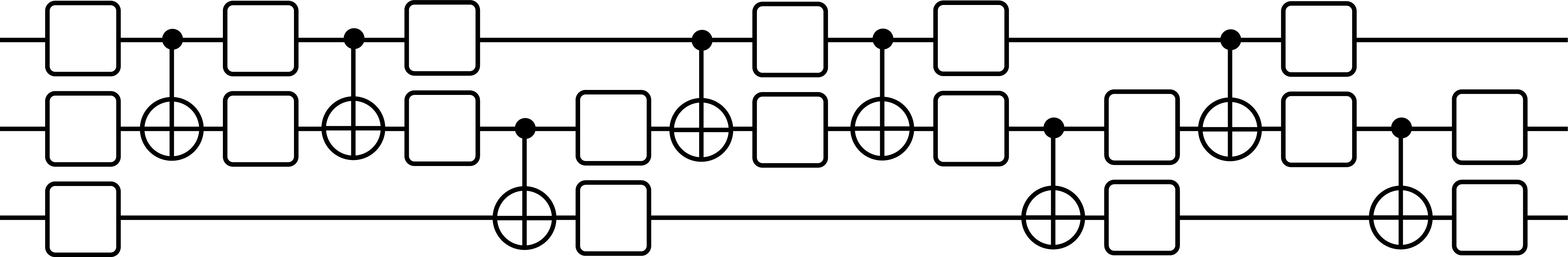}
        \caption{Template P}
        \label{fig:template_p}
    \end{subfigure}
    \caption{Labeled PQC templates that commonly appear when synthesizing unitaries. Labels in these figures correspond to the PQC Template Labels in Fig.~\ref{fig:histograms}. Squares represent parameterized U3 gates.}
    \label{fig:pqc_templates}
\end{figure}

\begin{figure}
    \centering
    \begin{subfigure}[b]{0.47\textwidth}
        \includegraphics[width=\textwidth]{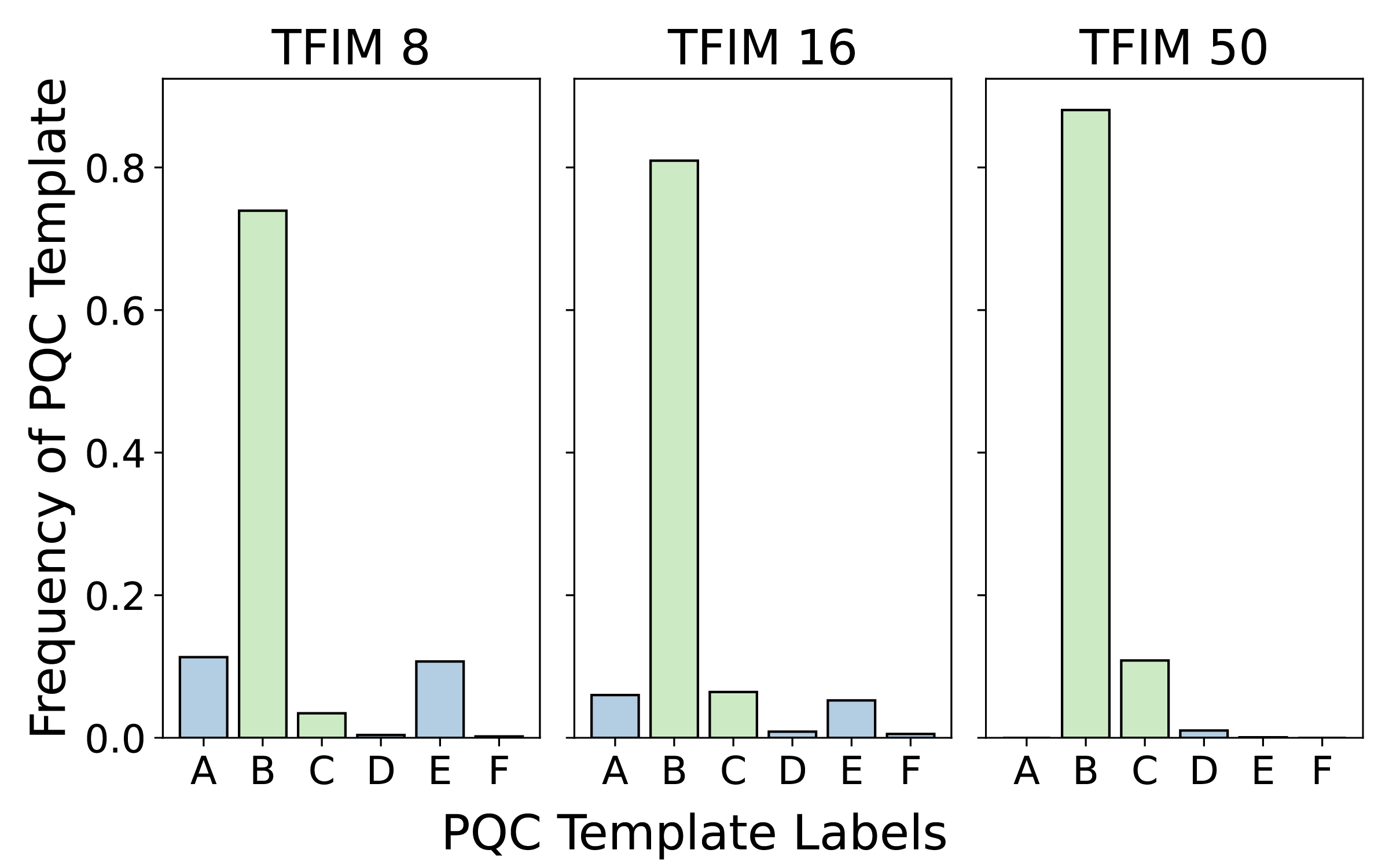}
        \caption{PQC templates that appear in synthesized TFIM circuits of various widths. Circuits of different widths share characteristic modes around the PQCs labeled B and C.}
        \label{fig:tfim}
    \end{subfigure}
    \par\bigskip
    \begin{subfigure}[b]{0.47\textwidth}
        \includegraphics[width=\textwidth]{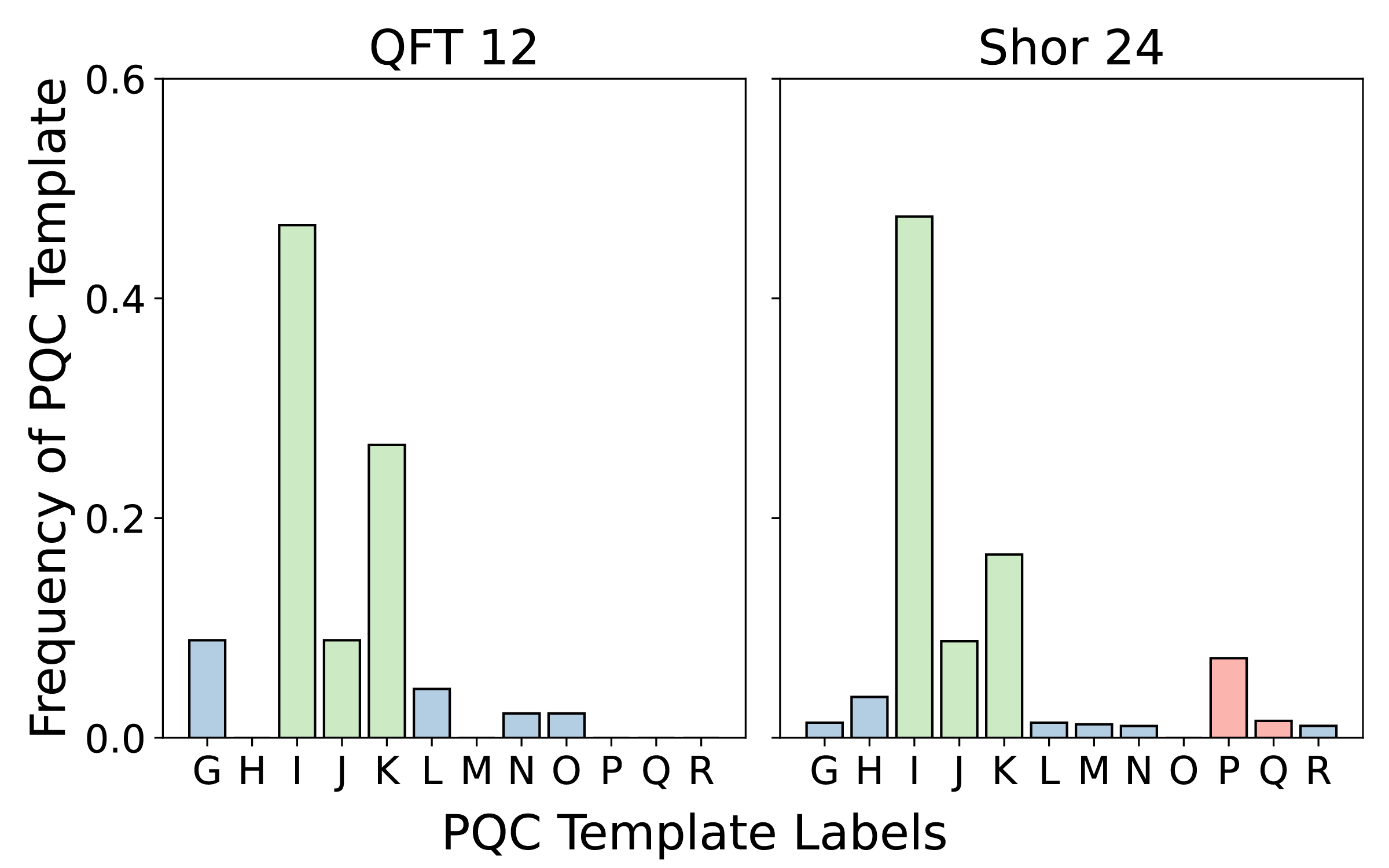}
        \caption{PQC templates that appear in synthesized QFT and modular exponentiation circuits. These related circuits have similar modes around templates I, J, and K. The more complex Shor's modular exponentiation circuit has extra templates not found in the QFT.}
        \label{fig:qft_shor}
    \end{subfigure}
    \par\bigskip
    \caption{Histograms of PQC templates that appear in the optimized TFIM circuits of various widths (\ref{fig:tfim}), and QFT and Shor modular exponentiation circuits (\ref{fig:qft_shor}).}
    \label{fig:histograms}
\end{figure}

\subsection{Common Structures in Partitioned Quantum Circuits}
Across many runs of synthesis, we have observed two patterns. First, partitioned unitaries from different width circuits of the same family tend to have similar characteristics. Second, unitaries from circuits that are related but not from the same family also tend to share characteristics. Critically, both of these points remain true even when the unitaries from these circuits are not identical. We can begin to measure this by examining what PQCs commonly appear once unitaries from various circuits are synthesized. Fig.~\ref{fig:pqc_templates} illustrates four PQCs that often implement partitioned unitaries.

Fig.~\ref{fig:tfim} illustrates the first point. Across multiple circuit widths, synthesized unitaries from the TFIM circuit family have shared structure. This figure shows the distribution of PQCs that implement unitaries from these TFIM circuits. Fig.~\ref{fig:qft_shor} illustrates the second point. In this case, the modular exponentiation ``Shor" and QFT circuits have replicated modes in their distributions of selected PQC templates. Unitaries taken from partitioned QFT and modular exponentiation circuits are similar, but not identical. We can also see from Fig.~\ref{fig:qft_shor} that the modular exponentiation circuit has added complexity compared to the QFT, as seen by the extra mode around the PQC labeled ``P" (Fig.~\ref{fig:template_p}).

We expect that a learned mapping might be advantageous in this scenario as there are clearly shared characteristics across related circuits. Because the unitaries in these circuits are not the same, memorizing unitary-PQC pairs is not helpful in predicting what structures might appear for new, unseen, unitaries. Training a model on a variety of circuit families and across many circuit widths aids us in learning a mapping that is robust to unseen unitaries. Section \ref{sec:training} further discusses the efforts we have made to highlight this phenomenon.

\section{QSeed}
\label{section:qseed}
\begin{figure}
    \centering
    \includegraphics[width=0.4\textwidth]{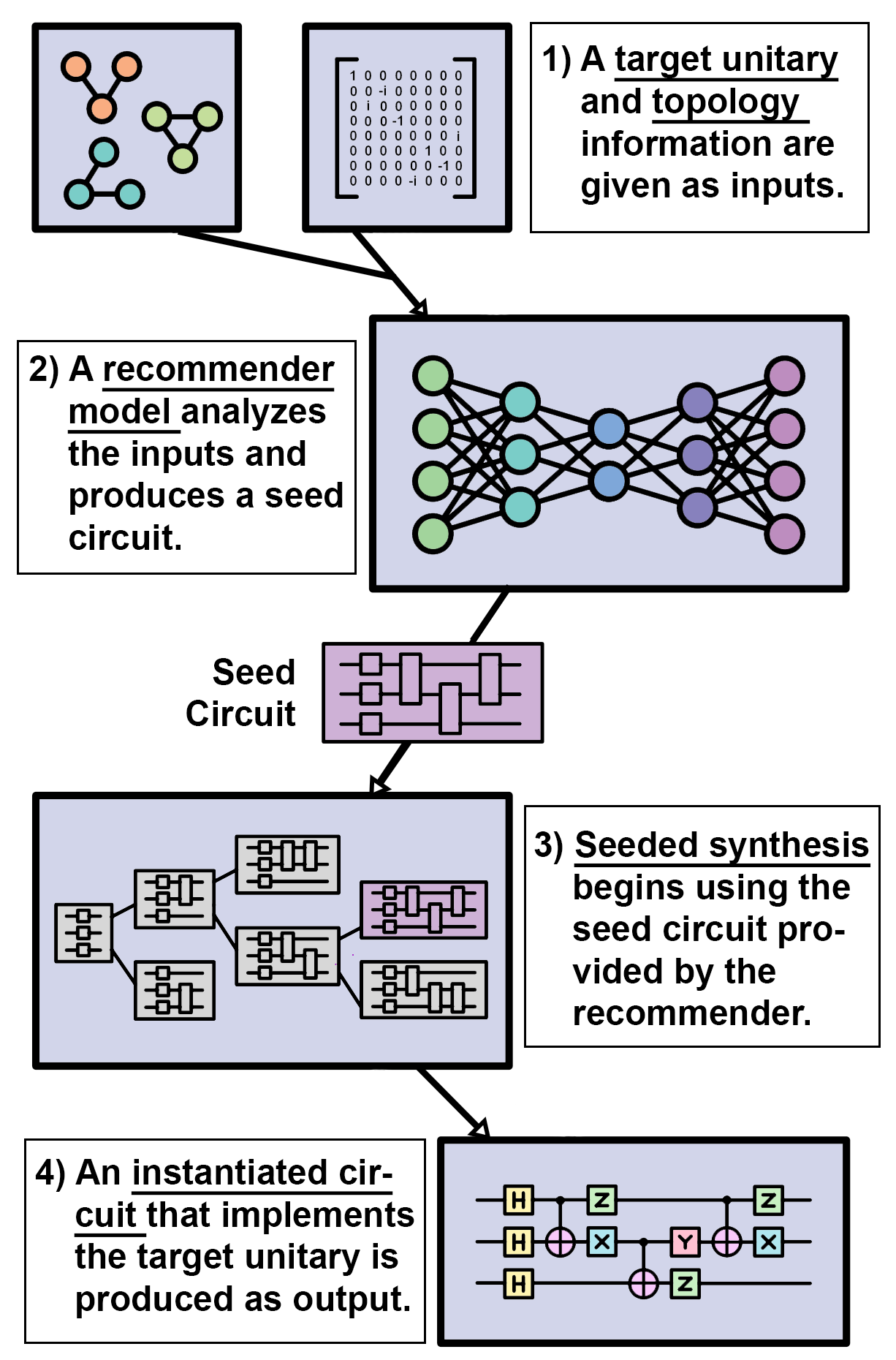}
    \caption{QSeed operational flow graph. QSeed takes as input a target unitary and topology information, and uses a recommender model to produce a seed PQC. This circuit is used as the starting point for seeded synthesis. The output is a circuit that implements the target unitary.}
    \label{fig:flow}
\end{figure}

This section presents QSeed, a unitary synthesis tool which consists of two major components:
\begin{enumerate}
    \item The recommender model: a deep  neural network which uses information about a target unitary to propose PQCs that are likely to instantiate the target unitary.
    \item Seeded synthesis: a unitary synthesis algorithm which begins the synthesis process at a provided seed circuit.
\end{enumerate}
The programmatic flow of the QSeed algorithm is depicted in Fig.~\ref{fig:flow}.

\subsection{The Recommender Model} \label{sec:recommender_model}
We have implemented a recommender model as a deep autoencoder neural network. This choice was made based off the observation that unitary matrices taken from partitioned quantum circuits exhibit low dimensionality. Section \ref{section:low_dim} empirically illustrates this observation by use of a PCA of unitaries taken from a wide variety of partitioned quantum circuits. Dimensionality reduction through the use of neural networks to learn low-dimensional manifolds is widely done in practice \cite{hinton_2006_reducingdimensionality}.

The three qubit recommender model takes as input a unitary matrix $U$, finds the canonical representation of the unitary (see Section \ref{section:canonical}), then flattens the real and complex elements of the unitary into a vector $x_{U} \in \mathbb{R}^{128}$. The autoencoder output is a vector $y_{U} \in \mathbb{R}^{|\mathcal{PQC}|}$. Each element in the model's output corresponds to a PQC in the set $\mathcal{PQC}(n,K)$. For $n=3$ qubit linear topologies and a maximum CNOT count of $K=8$, $|\mathcal{PQC}(n,K)| = 1199$. The value of $softmax(y_U)_i \in [0,1]$ represents the model's confidence that template $i$ can implement unitary $U$. 

During inference, the recommender model compresses the input unitary from an $\mathbb{R}^{128}$ to an $\mathbb{R}^{32}$ representation. Fig.~\ref{fig:pca} indicates that for our dataset of three qubit partitioned unitaries, this representation is capable of capturing at least 90.7\% of the variance within these matrices.

\subsection{PQC Templates} \label{sec:pqc_templates}
As mentioned in Section \ref{sec:recommender_model}, the output of the recommender model is a vector of size $1199$, each element corresponding to a possible output PQC template. These templates are generated by traversing the synthesis circuit tree in a breadth-first manner and enumerating each circuit. This recommender model was designed for the task of mapped circuit optimization, so only three qubit linear topologies are considered. Circuits with more than three consecutive two-qubit gates applied to the same qubits are not included, as three CNOTs are sufficient for implementing any two-qubit unitary \cite{tucci2005introduction}. All PQC templates conforming to these restrictions with zero to eight CNOT gates are considered. Example PQC templates are illustrated in Fig.~\ref{fig:histograms}.

\subsection{Training the Recommender}
\label{sec:training}
The recommender model is trained on a dataset of three qubit unitaries partitioned from twelve different circuit families. Each family of circuits contained multiple benchmarks of various widths. Each unitary in the dataset is paired with one PQC for each topology that is targeted. These ``label" PQCs are found by running synthesis and ensuring the unitary can be implemented by that PQC. 

To ensure generality across circuit widths, benchmarks from each circuit family were withheld from training. The test and training circuits are listed in Table \ref{tab:test_circuits}. Note that no Grover's algorithm circuits were included in the training dataset.

\begin{table}[]
    \centering
    \begin{tabular}{c|c|c}
        Circuit Name & Training Widths & Test Widths\\
        \hline
        add & 17, 65 & 41 \\
        grover & - & 10 \\
        heisenberg & 4, 6, 7, 8, 16, 32, 64 & 5 \\
        hhl & 8 & 6 \\
        hubbard & 4, 18, 50 & 8 \\
        mult & 8, 32, 64 & 16 \\
        qae & 11, 33, 101 & 65 \\
        qft & 3, 4, 8, 16, 32 & 64 \\
        qml & 4, 25, 60, 108 & 128 \\
        qpe & 6, 10, 14 & 18 \\
        shor & 16, 32 & 64 \\
        tfim & 3, 4, 5, 6, 7, 8, 16, 32, & 64 \\
        vqe & 12, 14 & 18 \\
    \end{tabular}
    \caption{Split of circuits withheld from recommender training for testing. No Grover's algorithm circuits are used in training.}
    \label{tab:test_circuits}
\end{table}
\begin{figure*}[h]
    \centering
    \includegraphics[width=0.98\textwidth]{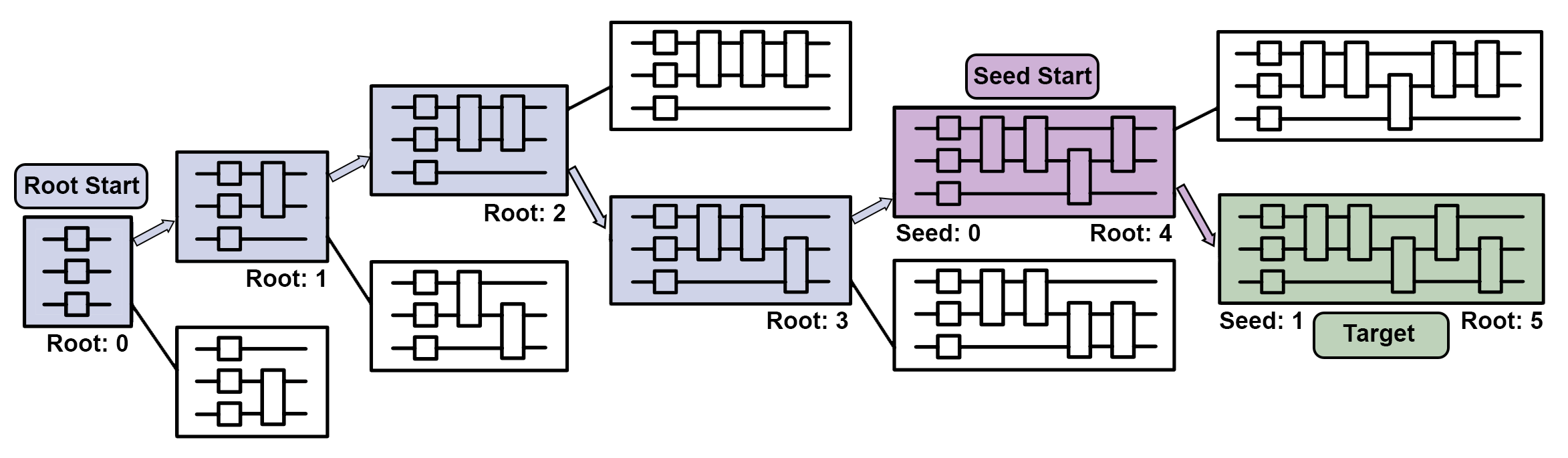}
    \caption{Seeded synthesis provides speed ups compared to synthesis which begins at the root because it allows for the circuit search process to start closer to the target. In this example, QSearch begins synthesis at the node labeled ``Root Start", and requires 5 hops (highlighted in blue) before reaching the target. Seeded synthesis instead begins at the node labeled ``Seed Start", and only requires 1 hop (highlighted in purple) before reaching the target.}
    \label{fig:search_tree}
\end{figure*}

The recommender model was pretrained as a denoising autoencoder with the task of reconstructing noisy input unitaries. Afterwards, a PQC template selection head is appended to the encoder portion of the network, and the network is then fine tuned to learn the mapping from unitaries to PQCs. The network must be conditioned on qubit topology information in order to propose seeds conforming to different connectivities. After training, the network was able to reach a top-3 classification accuracy of 60\%. This classification accuracy is deceptively low. As each unitary in the dataset is only associated with a single PQC label, even though multiple PQCs (often of the same depth) are capable of implementing a given unitary, good seed proposals may be classified as incorrect. Further tuning of the recommender network's architecture and hyperparameters for higher classification accuracy is surely possible.

\subsection{Seeded Synthesis}
As described in Section \ref{section:background}, there are two primary processes that happen during unitary synthesis: instantiation and tree traversal. This section describes how using seeded synthesis can alleviate the extensive run time of bottom-up synthesis by accelerating the tree traversal process.

Bottom-up synthesis algorithms, such as QSearch and LEAP, begin the circuit tree exploration process with a circuit that contains only single qubit rotation gates. This circuit is the root node of the circuit tree depicted in Fig.~\ref{fig:qsearch_tree}. Each node in the tree describes a different PQC that can be built by appending gates onto the root circuit.

It may be the case that some desired target circuit/node is situated very far down the tree. Fig.~\ref{fig:search_tree} provides an example of such a case. Here, the path from the ``Root Start" node to the ``Target" node has distance five. Each of the six circuits along this path must be instantiated before a solution is found. 

Seeded synthesis begins the circuit tree traversal at a non-root node. In Fig.~\ref{fig:search_tree}, the ``Seed Start" node is highlighted in purple. The path from this node to the target has distance one, and only two instantiation calls are made, one for each circuit along this path. When good seed circuits are provided, seeded synthesis can perform synthesis far faster than bottom-up synthesis, as fewer parameterized circuits must be instantiated. In practice, many deep parameterized circuits are very likely to be able to implement some target unitary. This means that proposing ``bad" seed circuits, which correspond to high gate count solutions, will still provide speed ups. This phenomenon is further explored in Section \ref{section:evaluation}.

In order to account for scenarios where a given seed circuit contains too many gates, synthesis is also capable of traversing up the circuit tree. This means that gates can also be removed to examine less deep candidate circuits. The tree search process advances both towards the root and towards leaf nodes.

\section{Evaluation}
\label{section:evaluation}
To highlight the effectiveness and demonstrate the use of QSeed, we have evaluated it on the task of wide quantum circuit optimization. There are several proposed methods for using unitary synthesis to optimize wide quantum circuits, including QGo \cite{wu2021qgo} and TopAS \cite{weiden_2022_topas}. In both cases, wide quantum circuits are partitioned into smaller subcircuits which are synthesized independently. We demonstrate how using QSeed can accelerate QGo style optimization of mapped quantum circuits, while maintaining low multi-qubit gate counts. A partition width of three qubits was used. All benchmarks were run on a 64 core AMD EPYC 7702P processor. The synthesis tools that are evaluated here are available through the \emph{BQSKit} toolkit \cite{bqskit}.

\begin{figure*}[h]
    \centering
    \includegraphics[width=\textwidth]{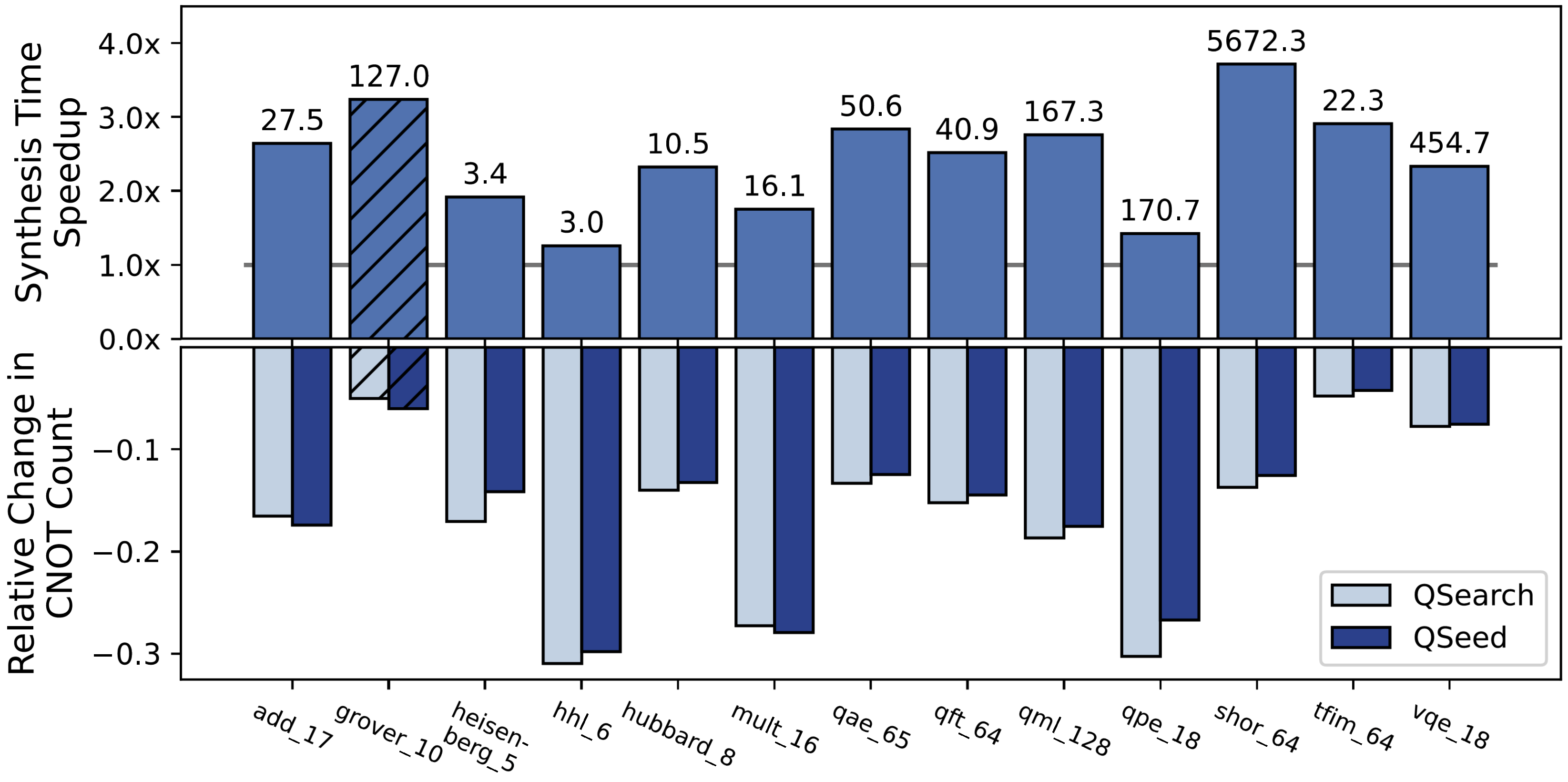}
    \caption{QSeed compared to QSearch on test circuits. The top graph shows speedup in wall-clock synthesis time and absolute time savings in seconds as data labels. Larger speedups are better. The bottom graph shows relative changes in CNOT gate count relative to Qiskit level 3 optimized circuits. Lower relative change in CNOT counts are better. Across all test circuits, QSeed has an average speedup of $2.4\times$, and an average of $1.0\%$ more CNOT gates compared to QSearch. Synthesis time includes the time needed for QSeed's recommender model do inference. These circuits were withheld from the training process (see Table \ref{tab:test_circuits}), and no Grover's algorithm circuits were included in the training dataset. These values represent an average of three trials.}
    \label{fig:main}
\end{figure*}

\subsection{Benchmark Generation}
For our training and test benchmark circuits, we generated a group of circuits that represent a wide breadth of important high level algorithms. These benchmarks can generally be grouped into variational algorithms, Quantum Fourier Transform (QFT) based algorithms, Quantum Machine Learning (QML) algorithms, and algorithms that show promise in the field of Quantum Finance. For our variational algorithms, we used Qiskit Nature to generate an ansatz for the ground state energy of molecular interaction \cite{qiskit}. We then simulated VQE to generate solved parameters for LiH, BeH2, BH3, and CH4 molecules. For QML, we first have an ZZ feature map to encode classical bit data into quantum data, then a two local network layer is applied. Our QFT based algorithms include standalone QFT, Shor's modular exponentiation, and general Quantum Phase Estimation (QPE) circuits. For quantum finance, we generated Quantum Amplitude Estimation (QAE) circuits which are commonly used for Monte Carlo Integration. Additionally, we generate Harrow-Hassidim-Lloyd (HHL) circuits applied to random square matrices as an example of a quantum linear solvers used in the field. \cite{herman_survey_2022}. Grover's algorithm circuits were implemented according to \cite{mandviwalla_2018_groversimplementation}. These circuits are able to be scaled to various widths and partitioned to fit into our quantum synthesis process.

\subsection{Compilation of Test Circuits} \label{sec:main_eval}
We evaluate the optimization of circuits based on two metrics: the time needed to perform synthesis and the number of multi-qubit gates in the optimized circuits. Each benchmark was optimized with Qiskit's level 3 transpiler optimization \cite{qiskit} and mapped to a square mesh qubit topology using the SABRE Swap algorithm \cite{li2019tackling}. After synthesizing with QSearch or QSeed, each block is also quickly optimized with scanning gate removal, as described in \cite{younis_2022_transpilation}.

Fig.~\ref{fig:main} compares QSeed and QSearch using these two metrics on the suite of test circuits outlined in Table \ref{tab:test_circuits}. On average, QSeed offers a speedup of $2.4\times$ compared to QSearch across all test circuits. The 64 qubit Shor's modular exponentiation circuit sees a speed up of $3.7\times$. In wall-clock time, this is a savings of nearly 5700 seconds. The 18 qubit VQE also notably sees a speedup of $2.3\times$. We have highlighted the VQE circuit results because due to the classical-quantum hybrid style of computation in variational algorithms, savings in synthesis time is of particular importance. These circuits offer large speedups due in part to their large sizes. There is a strong correlation between the number of partitions in a circuit and the amount of speedup compared to QSearch. On average, QSeed suffers a 1.0\% decrease in improvement in CNOT gate counts compared to QSearch. We feel that this difference is acceptable, and that with better recommender training techniques the gap between QSearch and QSeed optimized circuit gate counts can be closed.

The speedup of the 64 qubit Shor's modular exponentiation circuit is particularly pronounced. The reason for this speedup can be understood by observing the distribution of PQC structures that appear after synthesis. As shown in Fig.~\ref{fig:histograms}, the 24 qubit modular exponentiation circuit's PQC distribution contains a mode on the PQC template labeled ``P". As the width of the modular exponentiation circuit increases, this PQC becomes more prevalent. As it contains 8 CNOT gates, QSearch requires at minimum (and in practice, often more than) 9 instantiation calls to synthesize these partitions. QSeed is often able to do this in a single instantiation call. Circuits such as the the 6 qubit \emph{HHL}, and the 18 qubit \emph{QPE} circuits see comparatively little improvement in synthesis time. This is due to the fact that synthesizing partitioned unitaries from these circuits requires very few CNOTs. The cost of doing inference before synthesizing these unitaries diminishes potential speedups. The number of partitions, as well as the number of CNOTs in each partitions, but influence the amount of speedup QSeed delivers.

\subsection{Generalizing to Grover's}
No unitaries from Grover's algorithm circuits of any width were included in the training dataset for the recommender model. The compilation of the 10 qubit Grover's algorithm benchmark still results in a speedup of $3.2\times$, which has been emphasized with hatches in Fig.~\ref{fig:main}. This is a clear demonstration that QSeed is capable of offering large speedups in synthesis time even for completely unfamiliar circuits.

\subsection{Result Verification}
As outlined in Definition \ref{def:synthesis}, unitary synthesis allows for some tunable degree of approximation through the setting of the threshold constant $\epsilon$. It is therefore necessary that results be verified to ensure an acceptable amount of approximation error has accumulated in synthesized circuits. The results of \cite{patel_2022_quest} prove that the total accumulated error of a partitioned circuit is upper bounded by the sum of the individual approximation errors of each partition. To verify that the optimized circuits accurately implement the same functionality as the original circuits, we re-partition the optimized and original circuits into 12 qubit subcircuits, then sum the Hilbert-Schmidt distances of each corresponding 12 qubit block. The results of this experiment are summarized in Table \ref{tab:verification}. Each compiled test circuit accurately implements the desired functionality, and approximation errors remain many orders of magnitude lower than any error that might be introduced by noise.

\begin{table}[b]
    \centering
    \begin{tabular}{c|c}
         Circuit & Upper Bound on Approximation Error \\
         \hline
         add 41       & $1.4 \times 10^{-9}$ \\
         grover 10    & $2.4 \times 10^{-13}$ \\ 
         heisenberg 5 & $5.6 \times 10^{-15}$ \\
         hhl 6        & $2.2 \times 10^{-16}$ \\
         hubbard  8   & $3.1 \times 10^{-14}$ \\
         mult 16      & $2.5 \times 10^{-14}$ \\
         qae 65       & $3.6 \times 10^{-9}$ \\
         qft 64       & $2.4 \times 10^{-9}$ \\
         qml 128      & $4.2 \times 10^{-14}$ \\
         qpe 18       & $6.8 \times 10^{-13}$ \\
         shor 64      & $3.1 \times 10^{-7}$ \\
         tfim 64      & $1.8 \times 10^{-15}$ \\
         vqe 18       & $7.8 \times 10^{-16}$ \\
    \end{tabular}
    \caption{Cumulative error due to synthesis approximation in test circuits. The Shor benchmark has the largest amount of error as it contains the greatest number of partitions. Each optimized circuit accurately recreates the function of the corresponding original test circuit. }
    \label{tab:verification}
\end{table}

\subsection{Evaluating the Recommender}

Section \ref{sec:main_eval} empirically demonstrates the advantages of using QSeed for wide quantum circuit optimization. However, one can imagine other seed recommendation schemes to accelerate synthesis. For example, instead of using a network to produce seeds, one could simply randomly select seeds from a set of parameterized quantum circuits. 

Beyond measuring the time needed to perform synthesis, the speed of a seed selection strategy can also be compared by measuring the number of instantiation calls needed before a solution circuit is found. As mentioned in Section \ref{section:background}, instantiation must be performed for each candidate circuit in the synthesis process, and its complexity grows exponentially with the number of qubits in the circuit being synthesized. Synthesis is performed faster when fewer instantiation calls are made.

\begin{figure}[]
    \centering
    \includegraphics[width=0.49\textwidth]{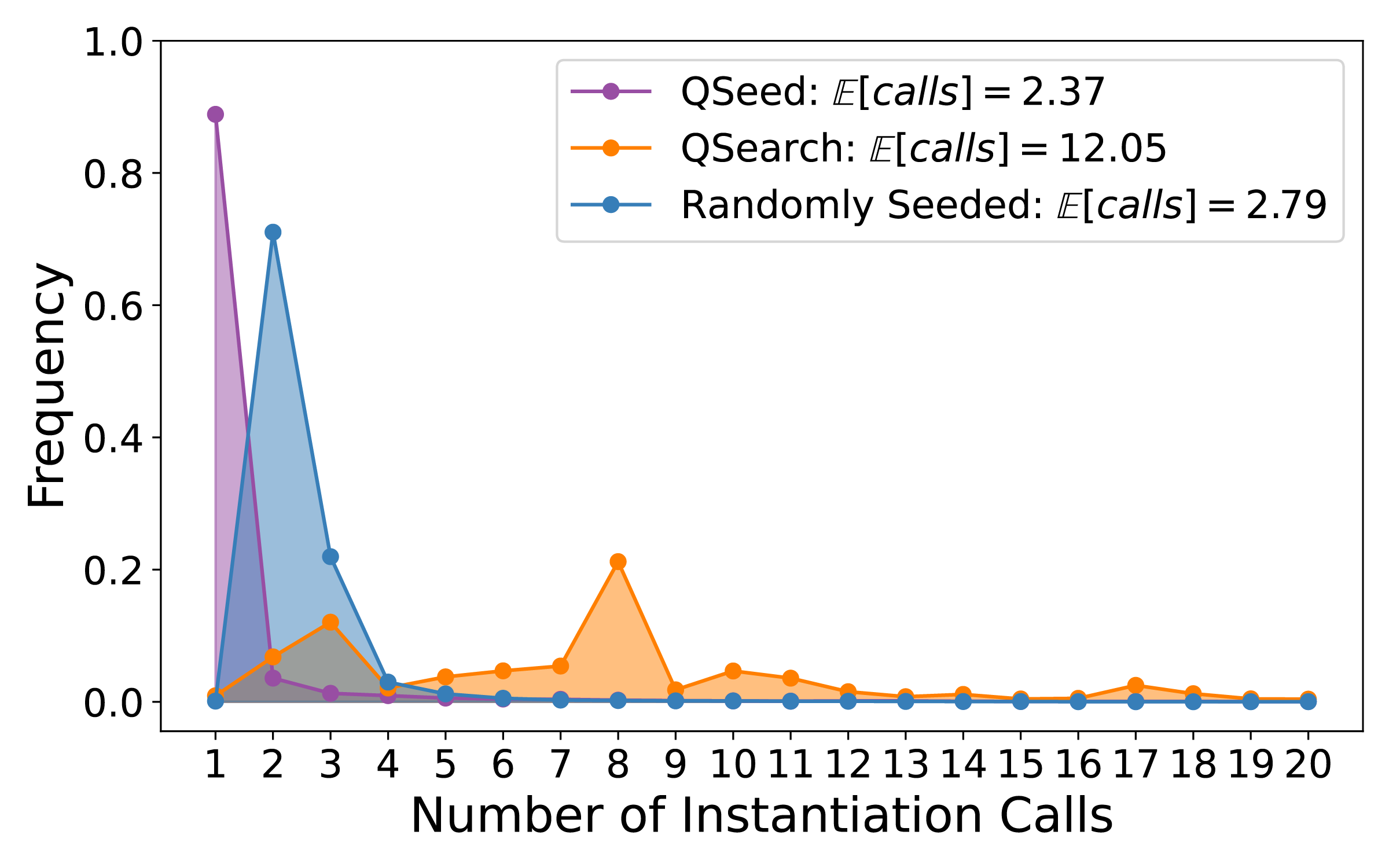}
    \caption{Distribution of instantiation function calls made. QSeed uses the recommender model while Randomly Seeded selects a random seed circuit for seeded synthesis. Both exhibit speed ups over QSearch because they decrease the expected number of instantiation calls. The X-axis is truncated at 20 instantiation calls, but continues up to 1704 for QSearch.}
    \label{fig:inst_calls}
\end{figure}

\begin{figure}[]
    \centering
    \includegraphics[width=0.49\textwidth]{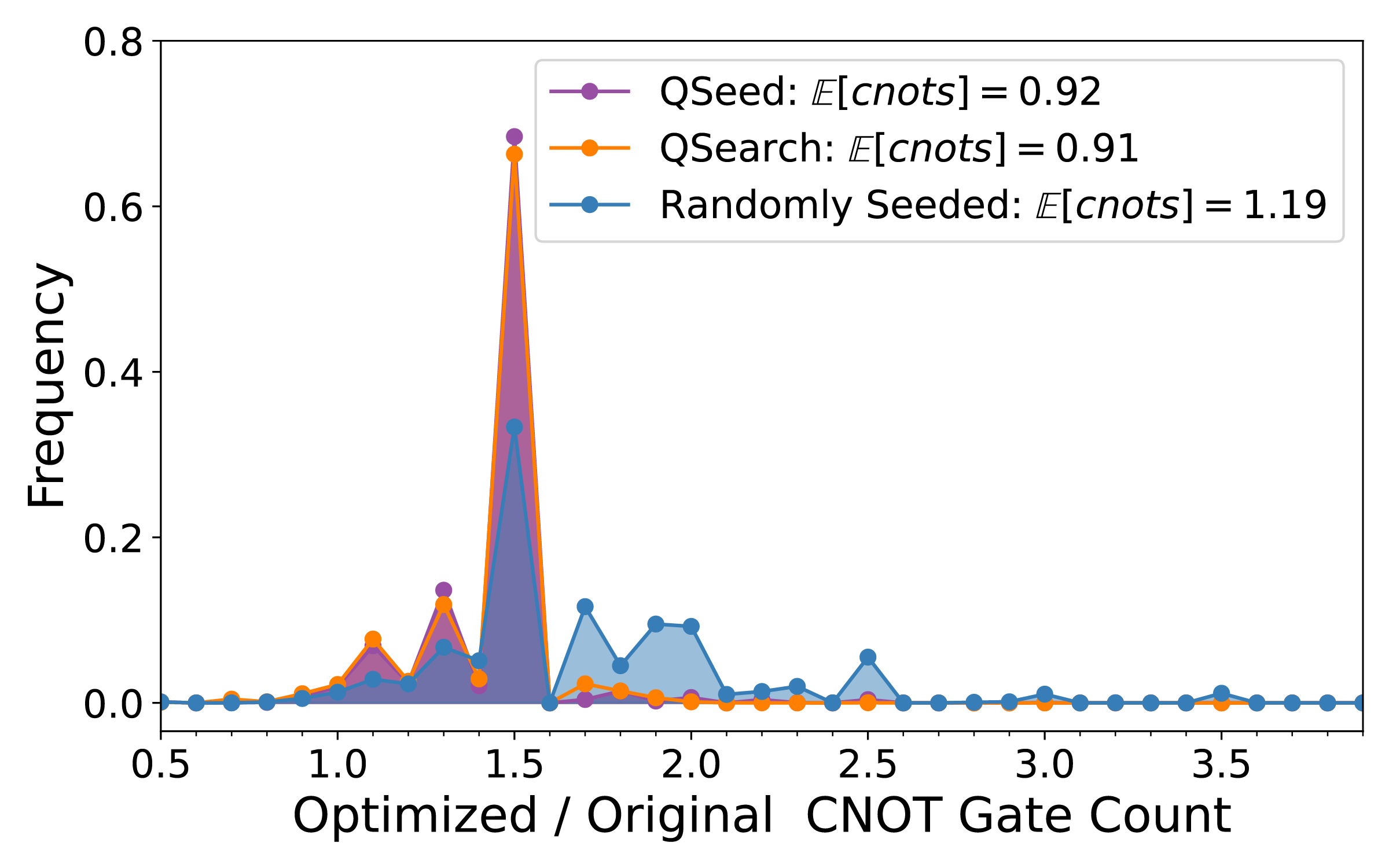}
    \caption{Distribution of relative CNOT gate counts. Values less than one correspond to fewer gates in the optimized compared to the original circuits (lower is better). The distribution of relative CNOT counts for QSeed closely matches that of QSearch. Randomly Seeded often results in circuit growth, which corresponds to values larger than 1.}
    \label{fig:cnots}
\end{figure}

Fig.~\ref{fig:inst_calls} compares the number of instantiation calls needed to synthesize blocks in the suite of benchmark circuits shown in Table \ref{tab:test_circuits} for three recommendation schemes. The first strategy is a random seed strategy, where three random circuits are sampled to act as synthesis seeds for each block. The second is the static QSearch strategy, where synthesis of each block begins at the circuit tree root as described in Section \ref{section:background}. The final strategy is the QSeed recommender model described in Section \ref{section:qseed}. Echoing the results of Fig.~\ref{fig:main}, QSearch tends to require more instantiation calls and takes more time to synthesize circuits. Similar to QSeed, the random initialization strategy tends to require very few instantiation calls. However, randomly selecting seeds degrades solution quality. Fig.~\ref{fig:cnots} illustrates the distribution of growth/shrinkage across all synthesized partitions. QSeed is able to match the quality of QSearch very closely. We can expect that both will decrease CNOT counts on average. The randomly seeded strategy tends to result in growth, as signified by the expected relative CNOT count of 1.19. An added benefit of seeded synthesis is that multiple seed locations can be considered simultaneously. Both the randomly seeded and QSeed recommender use this to their advantage, while QSearch must always begin synthesis at the same, single root seed.

As mentioned in Section \ref{sec:training}, the recommender model was trained until an accuracy of 60\% was reached on a random subset of unitaries from the test circuits (see Table \ref{tab:test_circuits}). Even so, Fig.~\ref{fig:inst_calls} demonstrates that most unitaries can be synthesized with a single instantiation call. This observation parallels the results of the random seed case, proposing any seed tends to lead to speedups compared to QSearch. The differences in CNOT gate counts highlighted in Fig.~\ref{fig:main} can therefore likely be diminished with a more accurate recommender model.

Overall, the results of Fig.~\ref{fig:inst_calls} and Fig.~\ref{fig:cnots} illustrate how QSeed is able to maintain speed advantages over QSearch, with minimal compromise in gate count.

\section{Discussion}
\label{section:discussion}
In this section, we provide further discussion relating to recommender generalization, the performance of QSeed, and the future of machine learning in quantum circuit optimization.

\subsection{Recommending General Unitaries}
This paper demonstrates that learning can be done for a subset of unitaries, unitaries taken from partitioned quantum circuits. How can this approach be made more generic and applied beyond the domain specific case? Our observations outlined in Section \ref{section:learning} detail how partitioned circuits can be characterized by distributions of PQC templates. The similarity between circuits of different domains and sizes can be evaluated using these distributions. We conjecture that across all algorithms expressed as quantum circuits, the prevalence of PQCs is not uniform. Certain templates are inherently more common than others. It is sufficient that we learn to associate unitaries with these important PQCs, rather than all PQCs. This explains why QSeed's recommender model is capable of generalizing to unseen circuits. With a well trained recommender model, it is possible that a few-shot learning approach, where only a small set of unitaries from unseen circuits are given, could further improve speedups and gate reductions. The results of Section \ref{section:evaluation}, especially the compilation of the completely unfamiliar 10 qubit Grover's circuit, demonstrate that learned mappings on this partitioned unitary dataset generalize to unseen partitioned unitaries.

The KAK decomposition demonstrates how to decompose arbitrary two-qubit unitaries \cite{tucci2005introduction}. Using ML to propose circuit templates differs in that it is not limited to two qubits, but it is not expected to work for general unitaries. A PCA of random unitaries shows that patterns are difficult to identify for the general (random) unitary case (Fig.~\ref{fig:pca}).

\subsection{Scaling Seeded Synthesis Width}
The results for this paper demonstrate a seeded synthesis algorithm for three qubit unitaries. Partitioning quantum circuits with different strategies and with larger partition widths still support the results illustrated in Fig.~\ref{fig:pca}. Unitaries of interest lie on low dimensional manifolds. Therefore, we expect that learning is still possible in these cases. The process of mapping unitaries to the PQCs that implement them becomes more challenging as the number of qubits increases. As discussed in Section \ref{section:background}, the number of gates needed to implement an arbitrary unitary of $n$ qubits grows with $O(4^n)$ complexity. If done in a topology-aware manner, more qubit topologies, and vertex labelings within these topologies must also be considered. Therefore, to apply QSeed beyond four or five qubits, more work must be done to narrow this space of possible output PQCs.

To this end, proposing circuit prefixes instead of PQCs that are expected to implement a unitary directly could still provide benefits. As illustrated in Fig.~\ref{fig:search_tree}, this corresponds to proposing PQCs that are simply closer to the target circuit than the root node.

As previously stated, we conjecture that a relatively small set of PQC templates are needed to implement all unitaries of interest. In the three qubit case, although there are 1199 possible PQCs to choose from (see Section \ref{sec:pqc_templates}), only about 415 appear for more than 10 instances in our dataset of over 203,000 unitaries. Of these, 285 appear more than 50 times, and 220 more than 100 times. For wider unitaries, it may therefore still be possible to enumerate all PQC templates that appear frequently.

The greatest complication with scaling this style of seeded synthesis to larger width unitaries is the availability (or lack thereof) of a large enough \emph{labeled} dataset of partitioned unitaries. The process of creating a labeled dataset for this style of optimization is outlined in Section \ref{sec:training}. Complications arise because the process of labeling unitaries with the PQCs that implement them is synthesis. Data augmentations that exploit knowledge of pre-labeled unitaries should therefore be explored.

\subsection{Future Directions}
Reinforcement Learning for circuit optimization has been explored in the past, notably in \cite{Fosel_2021}. We believe this approach remains a promising means of optimizing quantum circuits. The authors expect that training RL systems on datasets of non-random benchmarks can further advance real world performance of this style of peephole optimization. The use of RL as a method for searching the synthesis tree could also provide further speedups. Perhaps the use of the canonical representation of unitaries formulated in Section \ref{section:canonical} could aid in this venture.

Our presentation of a canonical representation of unitaries is also widely applicable to any domain where numerical methods are applied to unitary matrices. This canonical representation allows for the use of non-phase resilient distance metrics on unitaries, an area that is not currently widely explored.

As these unitary datasets are expensive to generate and label, it is also worth exploring possible data augmentations that preserve unitarity. We have preliminarily explored adding small amounts of random Gaussian noise to parameters to generate new unitaries that are closely related to existing unitaries (and can be implemented by the same PQC), but these experiments have not aided the recommender model's training.

As the size of unitaries used in ML tasks grow, using fully connected layers, as is done in the recommender model, will quickly become nonviable. It is unclear what neural network architecture is best suited for learning on unitary datasets. Convolutional neural networks are popular in computer vision tasks, as they are translationally equivarient and particularly adept at extracting spatially local patterns. We see no reason to believe that these inductive biases would be helpful for unitary datasets. Therefore, this domain deserves further attention.

\section{Conclusion}
\label{section:conclusion}
In this paper, we have demonstrated how machine learning can be applied directly onto unitary matrices to improve quantum circuit compilation and optimization. To our knowledge, we are the first to explore employing machine learning on unitary datasets for this purpose. We highlight how patterns in a diverse set of optimized real-world benchmarks emerge, and how these patterns imply learnable mappings between unitaries of interest and resource efficient circuits. We have introduced QSeed, a seeded unitary synthesis algorithm that replaces the expensive synthesis tree search process with machine learning inference. By mapping a canonical representation of unitaries to parameterized quantum circuits that may have generated them, the time needed to perform synthesis can be substantially improved with little change in the quality of solutions found. Notably, QSeed is able to synthesize a 64 qubit modular exponentiation circuit from Shor's algorithm $3.7\times$ faster than the state-of-the-art. Generalization of QSeed's recommender model, a deep neural network which proposes seed circuits, is demonstrated through the compilation of circuits not seen during training, including an entirely unfamiliar Grover's algorithm circuit. This project illustrates how simple machine learning techniques can make substantial improvements to quantum circuit compilation tools, and ensure their applicability to a wider scope of problems.

\section*{Acknowledgements}
\label{./section:acknowledgements}
This work was supported by the DOE under contract DE-5AC02-05CH11231 through the Office of Advanced Scientific Computing Research (ASCR) Quantum Algorithms Team and Accelerated Research in Quantum Computing programs, and by the NSF Challenge Institute for Quantum Computation (CIQC) program under award OMA-2016245.


\bibliographystyle{IEEEtranS}
\bibliography{refs}

\end{document}